\documentclass{article}
\usepackage{spconf,amsmath,graphicx}
\usepackage{graphicx}
\usepackage{amsfonts}
\usepackage{subcaption} 
\usepackage{float}     
\usepackage{booktabs}

\usepackage{stfloats}
\usepackage{url}

\title{OG-PCL: Efficient Sparse Point Cloud Processing for Human Activity Recognition}
\name{Jiuqi Yan$^{\dag,1}$, Yong Wang$^{1}$, Chendong Xu$^{1}$, Dongyu Liu$^{1}$  \thanks{ $^\dag$ Corresponding author.}}

\address{$^1$Key Laboratory of Underwater Acoustic Signal Processing of Ministry of Education,\\ Southeast University, Nanjing, 211189, China\\
}
\begin{document}
\ninept
\maketitle

\begin{abstract}  
Human activity recognition (HAR) with millimeter-wave (mmWave) radar offers a privacy-preserving and robust alternative to camera- and wearable-based approaches. In this work, we propose the Occupancy-Gated Parallel-CNN Bi-LSTM (OG-PCL) network to process sparse 3D radar point clouds produced by mmWave sensing. Designed for lightweight deployment, the parameter size of the proposed OG-PCL is  only 0.83M  and achieves 91.75\% accuracy on the RadHAR dataset, outperforming those existing baselines such as 2D CNN, PointNet, and 3D CNN methods.  We validate the advantages of the tri-view parallel structure in preserving spatial information across three dimensions while maintaining efficiency through ablation studies. We further introduce the Occupancy-Gated Convolution (OGConv) block and demonstrate the necessity of its occupancy compensation mechanism for handling sparse point clouds. The proposed OG-PCL thus offers a compact yet accurate framework for real-time radar-based HAR on lightweight platforms.  
\end{abstract}

\section{Introduction}
\label{sec:intro}

Non-contact sensing has emerged as a compelling alternative for human activity recognition (HAR), due to its potential for privacy preservation and robust performance under varying environmental conditions \cite{de2024real} \cite{arab2022convolutional}. Among non-contact modalities, millimeter-wave (mmWave) radar offers unique advantages: it can operate in non-visible environments, is less sensitive to lighting and occlusions, and provides rich motion cues through range–doppler–angle information and, increasingly, 3D point clouds  \cite{li2023radar} \cite{ullmann2023survey} \cite{xu2021review} \cite{camuffo2022recent} \cite{zhang2021comprehensive}. These properties make radar-based HAR attractive for privacy-sensitive and lightweight deployments \cite{ding2024milliflow} \cite{kong2023study} \cite{zhao2021human} \cite{liang2022fall}. However, existing methods often rely on dense representations or large-scale networks \cite{jiang2020millimeter}, which hinders real-time operation on resource-constrained devices.  

Most recent studies on radar-based HAR have primarily focused on dense point clouds produced by 4D millimeter-wave radars, where larger antenna counts provide finer spatial resolution \cite{ahmed2024advancements} \cite{jiang2020millimeter}. In contrast, sparse point clouds from cost-effective 3D radars are prevalent in lightweight scenarios but remain underexplored \cite{alhazmi2021machine}, partly due to the difficulty of robustly extracting informative features from sparse data. Moreover, many state-of-the-art approaches trade efficiency for accuracy, limiting real-world deployment \cite{xue2021mmmesh}. 

Singh et al. introduced the RadHAR dataset together with a time-distributed CNN–BiLSTM framework \cite{singh2019radhar}. While lightweight, their model collapses one spatial dimension of the voxelized point cloud, inevitably discarding structural information. Engel et al. proposed PntPoseAT, which integrates PointNet, LSTM, and self-attention calibrated by optical motion capture  \cite{engel2025advanced}. Although effective, this approach adopts a large parameter count and lacks adaptation to sparse radar data. Shi et al. leverage voxelized point cloud cubes with CNN–BiLSTM modules to bolster robustness against noise \cite{shi2023robust}; however, the reliance on 3D convolutions imposes significant computational overhead. Zhu et al. present mRadHPRS, applying PointNet++ and attention mechanisms directly to raw 3D point clouds  \cite{zhu2024mradhprs}. Despite achieving strong recognition accuracy, this design entails substantial model complexity without lightweight considerations. Beyond radar, transformer-based architectures for point-cloud modeling in computer vision \cite{guo2021pct} offer valuable insights, yet these methods typically assume dense sampling or large computational budgets, limiting their applicability to sparse, lightweight radar settings. This gap motivates the need for specialized architectures that explicitly address sparsity while maintaining computational efficiency.

To address these gaps, we propose the Occupancy-Gated Parallel-CNN Bi-LSTM (OG-PCL), a lightweight framework tailored for sparse radar point clouds. The network projects point clouds into three orthogonal views, enabling efficient spatial representation at low computational cost. Each view is processed by a view-specific CNN stack built from Occupancy-Gated Convolution (OGConv) blocks, inspired by partial convolutions \cite{liu2018image}, which explicitly handles sparsity by masking empty regions and applying a compensation mechanism. The per-view features are fused and modeled temporally by a Bi-LSTM \cite{zhang2015bidirectional}, yielding robust HAR performance with a small parameter budget.  


On the RadHAR dataset, OG-PCL achieves competitive accuracy with a compact model, highlighting its potential for real-time HAR on lightweight platforms. The main contributions of this work are summarized as follows:  
\begin{enumerate}  
  \item We design a tri-view projection strategy that preserves spatial information across three dimensions while maintaining low computational cost.  
  \item We introduce Occupancy-Gated Convolution (OGConv), a sparsity-aware module that adaptively modulates feature extraction in sparse regions and employs a compensation factor to stabilize responses.  
  \item We demonstrate that OG-PCL delivers strong accuracy with a very light parameter budget, making it suitable for lightweight deployment in radar-based HAR. 
\end{enumerate}  
The remainder of this paper is organized as follows. Section 2 describes the voxel data preprocessing and the OG-PCL architecture in detail. Section 3 presents the experimental setup, RadHAR dataset results, and ablation studies that quantify the contributions of the tri-view design and OGConv. Section 4 discusses related work and Section 5 concludes with directions for future research.  

\section{Methodology}
\label{sec:pagestyle}

\subsection{Voxel Data Preprocessing}

Millimeter-wave FMCW radars transmit chirp signals and estimate target reflections by applying a range–Doppler–angle processing pipeline. Specifically, raw ADC samples are transformed via range FFT, Doppler FFT, and angle-of-arrival estimation, followed by clutter removal and CFAR-based detection. This produces 3D point clouds in Cartesian coordinates $(x,y,z)$ with associated intensity values.  

Formally, each radar detection can be expressed as:
\begin{equation}
(x,y,z) = R \cdot 
\begin{bmatrix}
\cos\theta \cos\phi \\
\sin\theta \cos\phi \\
\sin\phi
\end{bmatrix}, 
\quad 
v = \frac{\Delta f_d \lambda}{2},
\end{equation}
where $R$ is the estimated range, $(\theta,\phi)$ are azimuth and elevation angles, $\Delta f_d$ is the Doppler frequency shift, and $\lambda$ is the radar wavelength.  

Following the RadHAR dataset \cite{singh2019radhar}, the resulting point clouds are voxelized into occupancy grids of size $10 \times 32 \times 32$ (depth $\times$ height $\times$ width), with each voxel storing the number of points falling into the corresponding spatial bin. Stacking $T=60$ frames yields a spatio-temporal voxel sequence $\mathbf{V} \in \mathbb{R}^{T \times X \times Y \times Z}$.  

To reduce the computational burden of 3D convolutions while preserving spatial information, each voxel frame $\mathbf{V}_t$ is projected along three orthogonal axes using max pooling:
\begin{align}
\mathbf{P}_t^{\mathrm{top}}(y,z) &= \max_{x}\mathbf{V}_t(x,y,z), \\
\mathbf{P}_t^{\mathrm{front}}(x,z) &= \max_{y}\mathbf{V}_t(x,y,z), \\
\mathbf{P}_t^{\mathrm{side}}(x,y) &= \max_{z}\mathbf{V}_t(x,y,z),
\end{align}
where $\mathbf{P}_t^{\mathrm{top}}$, $\mathbf{P}_t^{\mathrm{front}}$, and $\mathbf{P}_t^{\mathrm{side}}$ denote the top, front, and side projections, respectively. These 2D representations serve as the actual inputs to OG-PCL. 

\subsection{Occupancy-Gated Parallel-CNN Bi-LSTM Network (OG-PCL)}

\begin{figure*}[ht]
    \centering
    \includegraphics[scale=0.305]{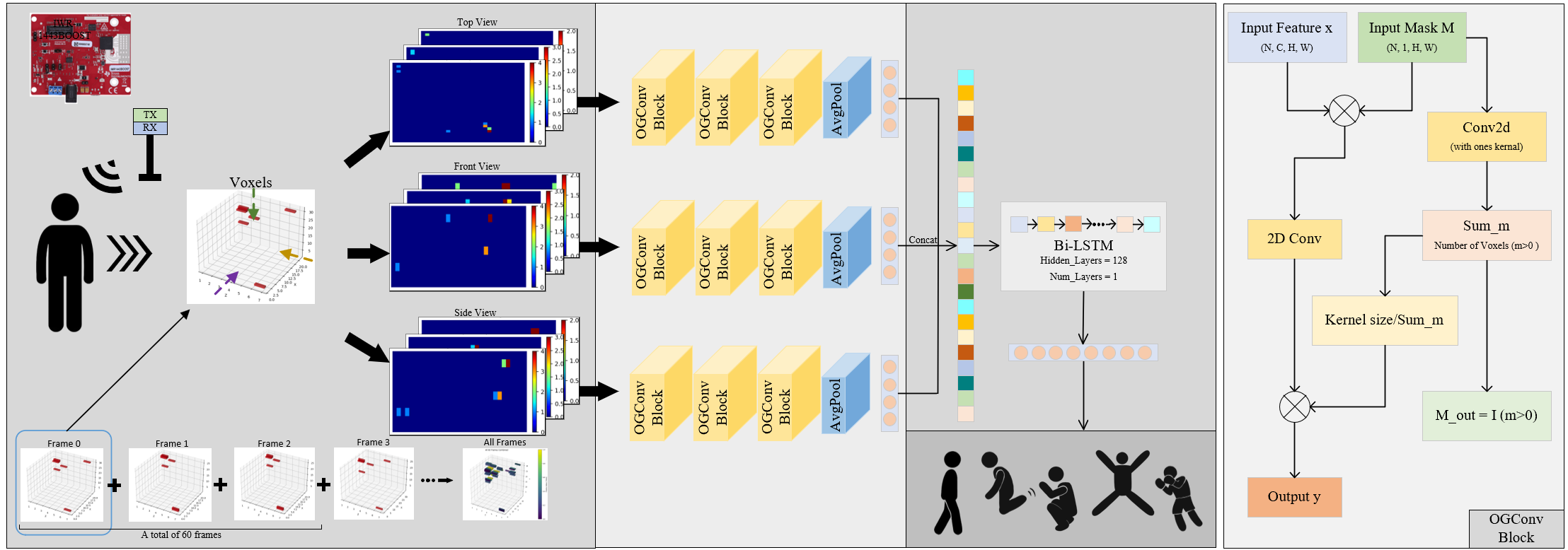}
    \caption{Structure of Occupancy-Gated Parallel-CNN Bi-LSTM Network and OGConv Block.}
    \label{OG-PCL}
\end{figure*}

The overall structure of Occupancy-Gated Parallel-CNN Bi-LSTM (OG-PCL) and the details of the Occupancy-Gated Convolution (OGConv) module are shown in the Fig.~\ref{OG-PCL}. We will introduce the principles of the parallel structure and OGConv block.

\subsubsection{Network overview and per-frame feature aggregation}

The three branches of OG-PCL can simultaneously process projection maps in three directions, to obtain more comprehensive dimensional information. The convolutional structure of each branch is two-dimensional, significantly reducing the number of parameters. 

Each view $v\in\{\mathrm{top,front,side}\}$ is processed by a view-specific CNN stack built from OGConv blocks (OGConv + BN + ReLU + AvgPooling). For the $t$-th frame the convolutional stack for view $v$ produces an intermediate spatial feature map, which we spatially aggregate to a per-frame vector via global pooling and a linear projection:
\begin{equation}
\mathbf{f}_v^{(t)} = \psi_v\!\big(\mathbf{P}_t^{v},\,\tilde{\mathbf{M}}_v^{(t)}\big)\in\mathbb{R}^{d_v},
\end{equation}
where $\psi_v(\cdot)$ denotes the view-specific OGConv-based backbone followed by adaptive average pooling and a fully-connected layer. $\tilde{\mathbf{M}}_v^{(t)}$ denotes the binary occupancy mask of frame $t$ under view $v$. The three per-view vectors are concatenated:
\begin{equation}
\mathbf{f}^{(t)} = \mathrm{Concat}\big(\mathbf{f}_{\mathrm{top}}^{(t)},\mathbf{f}_{\mathrm{front}}^{(t)},\mathbf{f}_{\mathrm{side}}^{(t)}\big)\in\mathbb{R}^{d},
\end{equation}
and the resulting sequence $\{\mathbf{f}^{(t)}\}_{t=1}^{T}$ is fed into a shared bidirectional LSTM for temporal modeling. Concat refers to channel-wise concatenation of features. $\mathbf{h}_t$ represents the BiLSTM output at time $t$; in our implementation we use the last-time-step output $\mathbf{h}_T$ for classification: 
\begin{equation}
\hat{\mathbf{y}} = \mathrm{Softmax}\big(\mathbf{W}\,\mathbf{h}_T + \mathbf{b}\big),
\end{equation}
where $\mathbf{W}$ and $\mathbf{b}$ denote the weights and bias of the final classification layer, Softmax is the standard normalization function to produce probability distributions. 

\subsubsection{Occupancy-Gated Convolution (OGConv)}

To handle the sparsity of voxelized point cloud data, we design OGConv to adaptively modulate convolution operations according to the occupancy status of each voxel. 

Given an input feature map $\mathbf{X} \in \mathbb{R}^{N \times C_\text{in} \times H \times W}$ and a binary occupancy mask $\mathbf{M} \in \{0,1\}^{N \times 1 \times H \times W}$, OGConv first performs an element-wise masking:
\begin{equation}
    \mathbf{X}_{m} = \mathbf{X} \odot \mathbf{M},
\end{equation}
where $\odot$ denotes Hadamard product, suppressing inactive regions. The element-wise masking suppresses inactive regions and ensures that only valid voxels contribute to convolution, while simultaneously providing the binary input for computing occupancy counts.  

The masked feature map $\mathbf{X}_{m}$ is then convolved with kernel $\mathbf{W} \in \mathbb{R}^{C_\text{out} \times C_\text{in} \times k_h \times k_w}$: 
\begin{equation}
    \mathbf{Y}_{\text{raw}} = \mathrm{Conv}(\mathbf{X}_{m}; \mathbf{W}),
\end{equation}
while the effective occupancy count for each local receptive field is computed by:
\begin{equation}
    D = \mathrm{Conv}(\mathbf{M}; \mathbf{1}),
\end{equation}
where $\mathbf{1}$ represents a convolution kernel of ones with shape $1 \times 1 \times k_h \times k_w$. 

To compensate for the loss of valid information due to sparsity, the convolution output is normalized by a scaling factor:
\begin{equation}
    \mathbf{Y} =
    \begin{cases}
        \mathbf{Y}_{\text{raw}} \cdot \frac{K}{D}, & D > 0,\\[6pt]
        0, & D = 0,
    \end{cases}
\end{equation}
where $K = k_h \times k_w$ is the kernel size. The scaling factor $K/D$ restores the magnitude of convolution outputs to be comparable with the dense case, preventing attenuation in sparse regions and stabilizing feature statistics. $K/D$ can also be regarded as a compensation factor. 

Finally, the output mask is updated to indicate valid activation regions for subsequent layers:
\begin{equation}
    \mathbf{M}_{\text{out}} = \mathbb{I}(D \geq 0),
\end{equation}
where $\mathbb{I}(\cdot)$ denotes the indicator function:
\begin{equation}
    \mathbb{I}(D \geq 0) =
    \begin{cases}
        1, & \text{if } D > 0, \\
        0, & \text{if } D = 0.
    \end{cases}
\end{equation}

Setting the output to zero when $D=0$ avoids propagating spurious noise in empty neighborhoods. The updated mask $\mathbf{M}_{\text{out}}$ explicitly indicates valid activation regions for subsequent layers, preserving semantic consistency across the network. 

\section{Implementation and Experiment Results}

\subsection{Data Preprocessing and Training Details}
Experiments were conducted on the open-source RadHAR dataset  \cite{singh2019radhar}, which was collected using a TI IWR1443BOOST FMCW radar operating in the 76–81 GHz band. Each sequence consists of T = 60 frames recorded at approximately 30 fps, with each frame represented as a voxelized 3D occupancy grid of size  $10 \times 32 \times 32$ (depth $\times$ height $\times$ width). The dataset covers five daily actions (Boxing, Jumping Jacks, Jumping, Squats, and Walking), performed by multiple participants under indoor conditions, with 39, 38, 37, 39, and 47 sequences per class, respectively. 

For efficient processing, voxel frames are projected into three orthogonal 2D views via max pooling: a top view ($32 \times 32$), a front view ($10 \times 32$), and a side view ($10 \times 32$). These projections serve as input to OG-PCL. Data augmentation, including randomized noise injection and geometric perturbations, was applied to improve generalization.

The model was trained using 5-fold cross-validation, with Adam (learning rate $1\times10^{-4}$), a batch size of 16, and cross-entropy loss for 50 epochs per fold. Training was performed on an NVIDIA A100 GPU. To ensure reproducibility, we provide the full implementation, including voxelization, tri-view projection, and 5-fold training scripts, at \url{https://github.com/BlakeYan97/OG-PCL}.

\subsection{Experiment Results and Analysis}

The proposed OG-PCL achieved a final accuracy of $91.75\%$ on RadHAR with only $0.83$ million parameters. To validate the advantages of OG-PCL, we compared it with representative baselines. The results are presented in Table~\ref{tab:model_comparison}. OG-PCL surpasses the RadHAR baseline (TD-CNN + LSTM) by more than $5\%$ in accuracy while remaining lightweight, and consistently outperforms 3D CNNs and PointNet with fewer parameters. These results confirm that the tri-view projection effectively preserves spatial occupancy cues with low computational cost. Although OG-PCL is $1.03\%$ less accurate than the Transformer-based PCT model, its significantly smaller parameter count ($0.83$M vs. $2.08$M) makes it better suited for deployment in resource-constrained scenarios. The combination of parallel multi-view design and OGConv contributes to robust performance under severe sparsity while keeping complexity minimal.  

\begin{table}[ht]
\centering
\caption{Models Comparison on RadHAR Dataset.}
\label{tab:model_comparison}
\resizebox{\columnwidth}{!}{%
\begin{tabular}{l|rrrr|r}

\hline Model & Acc. & Pre. & Rec. & F1. & Params(M) \\ \hline

MobileNet-V2 + LSTM & 0.8138 & 0.8320 & 0.8136 & 0.8140 & 3.67 \\ 
EffNet-B0 + LSTM & 0.8275 & 0.8419 & 0.8283 & 0.8287 & 5.45 \\ 
TD-CNN + LSTM \cite{singh2019radhar} & 0.8657 & 0.8790 & 0.8665 & 0.8674 & 0.31 \\ 
DenseNet-121 + LSTM & 0.8909 & 0.9050 & 0.8925 & 0.8921 & 8.16 \\ 
ResNet-18 + LSTM & 0.8915 & 0.9064 & 0.8922 & 0.8931 & 11.86 \\ \hline
3D CNN + LSTM & 0.9001 & 0.9134 & 0.9018 & 0.9017 & 10.67 \\ \hline
PointNet + LSTM & 0.9009 & 0.9065 & 0.9023 & 0.9029 & 3.93 \\ 
PCT + LSTM \cite{guo2021pct} & 0.9278 & 0.9356 & 0.9287 & 0.9305 & 2.08 \\ \hline
OG-PCL & 0.9175 & 0.9312 & 0.9187 & 0.9186 & 0.83 \\ \hline

\end{tabular}}
\end{table}

\begin{figure*}[ht]
\centering
\begin{subfigure}[b]{0.33\textwidth}
    \centering
    \includegraphics[width=\linewidth]{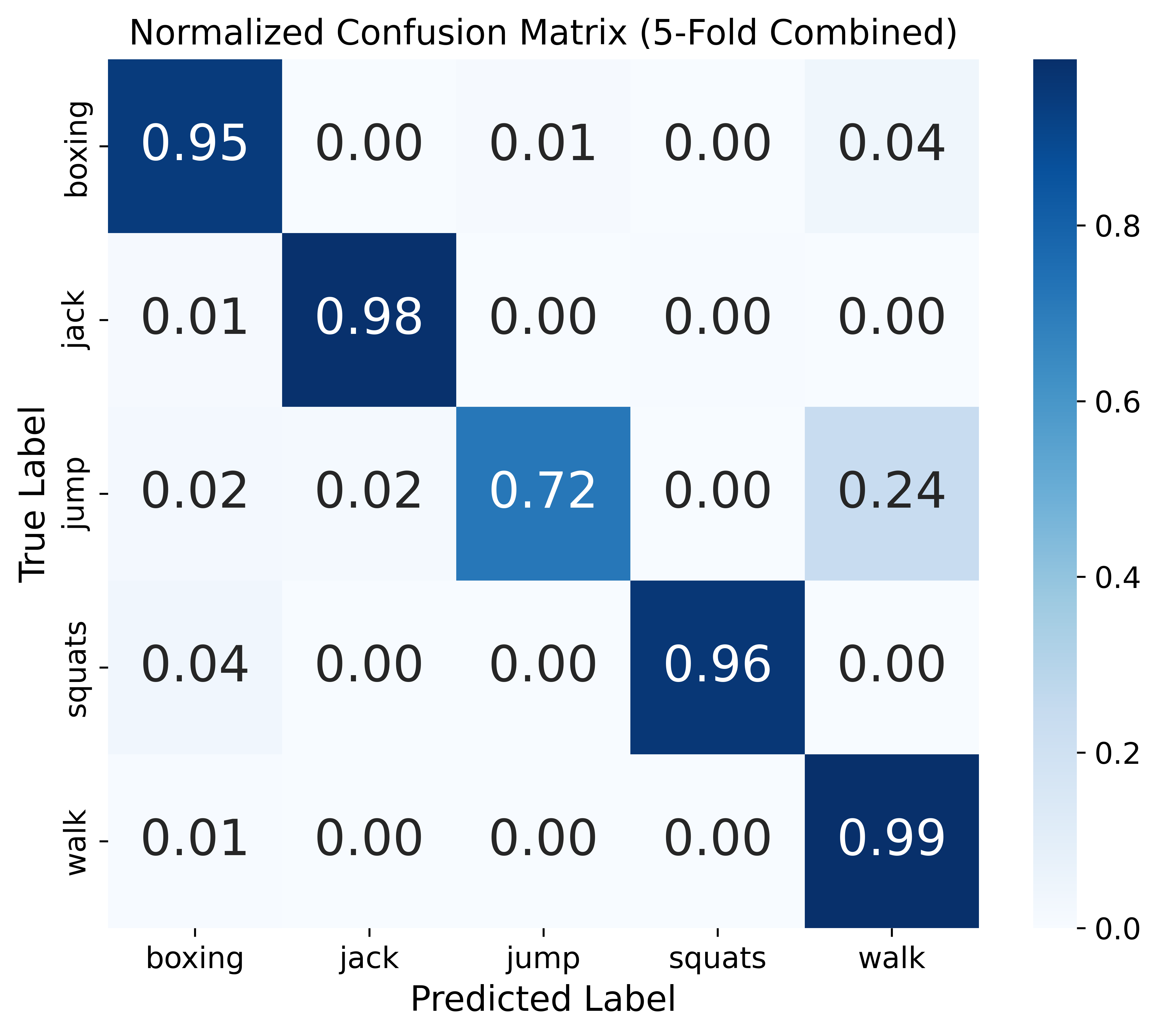}
    \caption{Confusion Matrix}
    \label{subfig:confusion_matrix}
    
\end{subfigure}
\hfill
\begin{subfigure}[b]{0.33\textwidth}
    \centering
    \includegraphics[width=\linewidth]{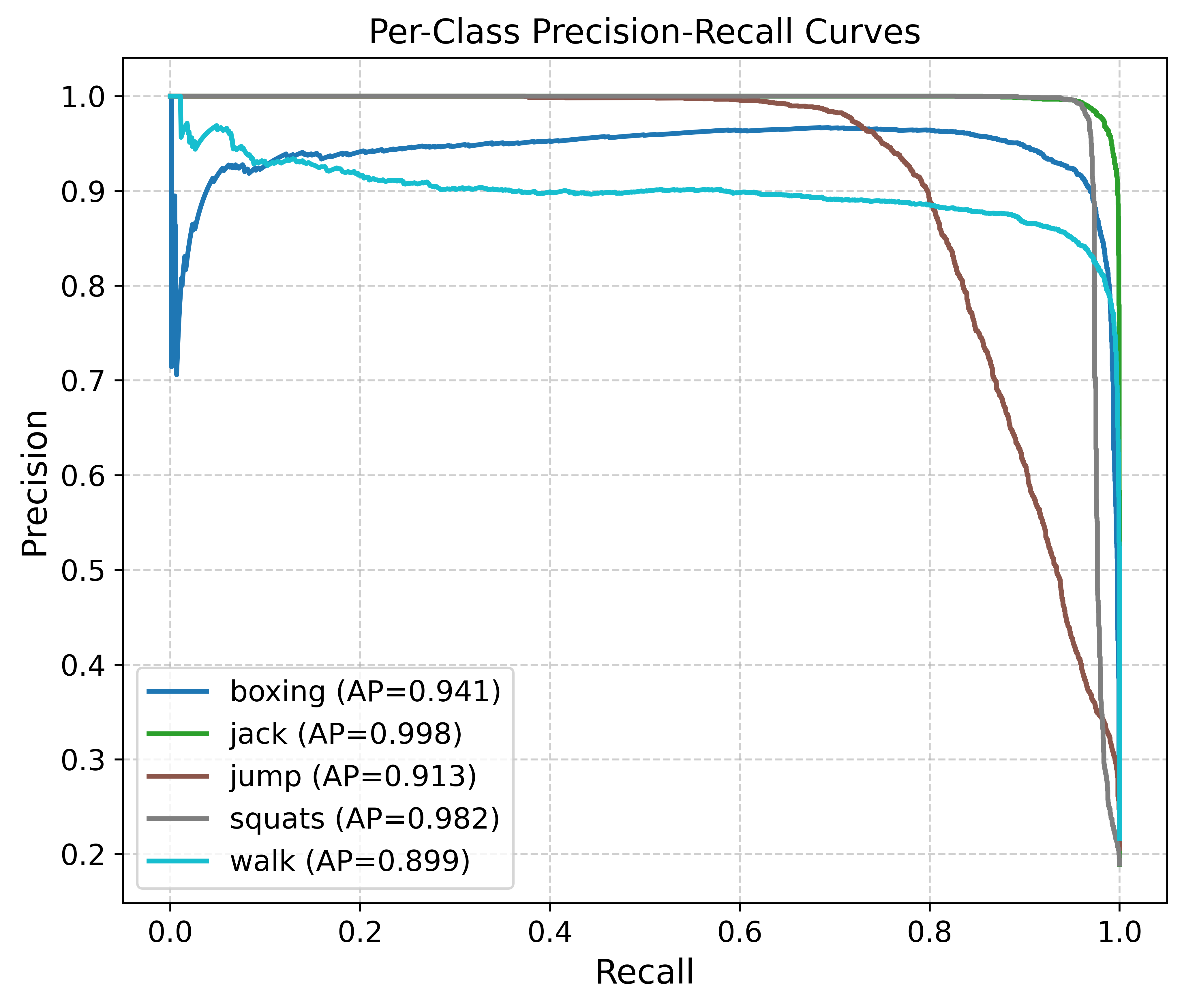}
    \caption{Precision-Recall Curve}
    \label{subfig:pr_curve}
\end{subfigure}
\hfill
\begin{subfigure}[b]{0.33\textwidth}
    \centering
    \includegraphics[width=\linewidth]{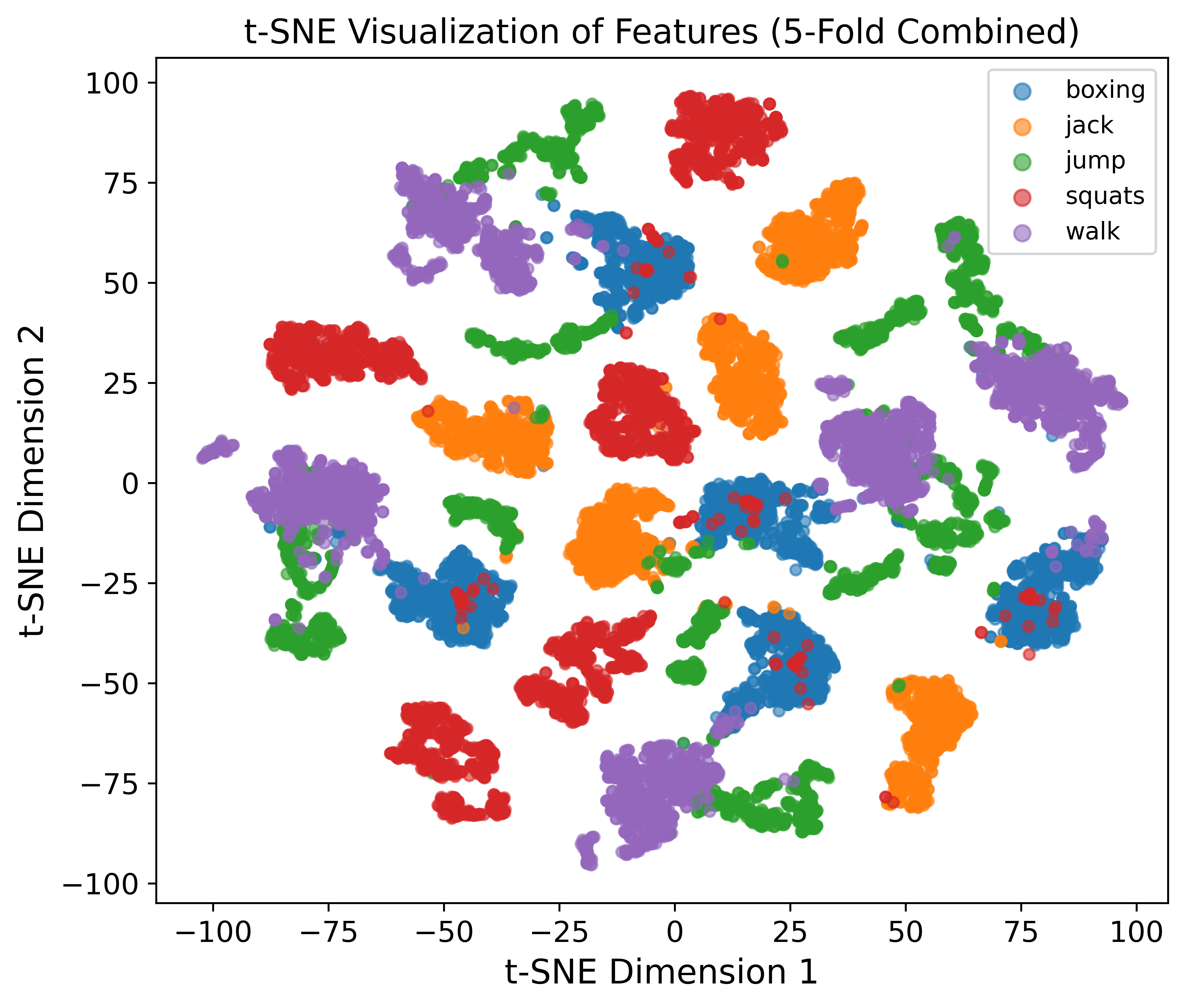}
    \caption{t-SNE Visualization}
    \label{subfig:tsne}
\end{subfigure}
\caption{Experimental results visualization: (a) Confusion Matrix of OG-PCL; (b) Per-Class Precision-Recall Curve; (c) t-SNE visualization.}
\label{fig:all_results}
\end{figure*}

Per-class results are summarized in Table~\ref{Per-Class Precision, Recall, F1-Score}. Most actions are recognized with high precision and recall, with the main confusions arising between \emph{Jump} and \emph{Walk} due to their similar motion patterns. Nevertheless, \emph{Jump} achieves a precision of $0.98$, while \emph{Walk} attains the highest recall of $0.99$. Figure~\ref{fig:all_results} further illustrates these patterns. The per-class PR curves (Fig.~\ref{subfig:pr_curve}) show that \emph{Jump} and \emph{Walk} have lower AP scores ($0.913$ and $0.899$, respectively) compared with other classes. The \emph{Jump} curve stays flat at precision $\approx 1.0$ until recall reaches $\sim0.7$, then drops sharply, indicating high confidence but limited coverage. In contrast, the \emph{Walk} curve decreases gradually from precision $1.0$ to $0.7$ over recall $[0,0.8]$, consistently below other classes, before surpassing \emph{Jump} beyond recall $0.8$, reflecting broader coverage at the expense of precision. These complementary trends explain the misclassifications concentrated in the \emph{Jump}–\emph{Walk} pair. The t-SNE visualization (Fig.~\ref{subfig:tsne}) also reveals stronger overlap between \emph{Jump} (green) and \emph{Walk} (purple) embeddings than among other categories. Finally, the confusion matrix (Fig.~\ref{subfig:confusion_matrix}) provides a concise summary of these inter-class confusions.

\begin{table}[ht]
\centering
\caption{Per-class accuracy, precision, recall, and F1-score on the RadHAR dataset.}
\label{Per-Class Precision, Recall, F1-Score}
\begin{tabular}{l|rrrrr}
\hline
       & Accuracy & Precision & Recall & F1-Score \\ \hline
Boxing & 0.95     & 0.92      & 0.95   & 0.94     \\ 
Jack   & 0.98     & 0.97      & 0.98   & 0.98     \\ 
Jump   & 0.72     & 0.98      & 0.72   & 0.83     \\ 
Squats & 0.96     & 0.99      & 0.96   & 0.97     \\ 
Walk   & 0.99     & 0.79      & 0.99   & 0.88     \\ \hline
\end{tabular}
\end{table}

\subsection{Ablation Experiments}

To assess the contributions of different components in OG-PCL, we conducted ablation studies focusing on the parallel architecture and the OGConv module.  

\subsubsection{Parallel Structures and Heterogeneous Branches}

We first evaluated the role of the parallel design by comparing single-branch CNN–BiLSTM models against the full parallel CNN–BiLSTM (PCL). Results in Table~\ref{tab:Parallel Structure and Hetero} show that the parallel structure improves accuracy by $1.97\%$–$5.12\%$ compared to individual branches, while increasing parameters only modestly (by $0.47$M). This demonstrates that combining complementary spatial views leads to stronger feature representations.  

To further optimize efficiency, we designed PCL (Hetero), where each branch is tailored to its projection resolution using heterogeneous convolutional configurations. As shown in Table~\ref{tab:Parallel Structure and Hetero}, PCL (Hetero) achieves a $0.8\%$ gain in accuracy over standard PCL while reducing parameters by $0.11$M, indicating that heterogeneous branch design provides a better trade-off between accuracy and efficiency.  



\begin{table}[ht]
\centering
\caption{Ablation Experiment of Parallel Stucture and Heterogeneous Design.}
\label{tab:Parallel Structure and Hetero}
\resizebox{\columnwidth}{!}{%
\begin{tabular}{l|rrrr|r}

\hline Model & Acc. & Pre. & Rec. & F1. & Params(M) \\ \hline

SingleBranch\_top & 0.8437 & 0.8681 & 0.8454 & 0.8446 & 0.41 \\
SingleBranch\_front & 0.8739 & 0.8860 & 0.8766 & 0.8717 & 0.41 \\
SingleBranch\_side & 0.8752 & 0.9003 & 0.8769 & 0.8775 & 0.41 \\ \hline
PCL & 0.8949 & 0.9081 & 0.8967 & 0.8946 & 0.88 \\
PCL (Hetero) & 0.9029 & 0.9130 & 0.9037 & 0.9043 & 0.77 \\ \hline

\end{tabular}}
\end{table}

\subsubsection{Occupancy-Gated Convolution}



We next assessed the impact of OGConv by replacing standard convolutions in PCL (Hetero). As shown in Table~\ref{tab:OGConv}, OG-PCL improves accuracy from $90.29\%$ to $91.75\%$ with only $0.06$M additional parameters. This highlights OGConv’s ability to preserve informative features under sparse voxel inputs.  

To further isolate the role of the compensation factor $K/D$, we removed it from OGConv. Accuracy dropped to $90.54\%$, confirming that rescaling is critical for mitigating feature attenuation in sparse receptive fields. Interestingly, without $K/D$, precision is relatively high ($0.9209$) but recall decreases, indicating that the model becomes more conservative and fails to capture some true positives. Restoring the compensation factor balances precision and recall, yielding the highest F1-score ($0.9186$).  

Together, these results demonstrate that both the parallel multi-view design and the sparsity-aware OGConv module are essential to OG-PCL’s performance, with each component contributing complementary gains in accuracy and efficiency.  

\begin{table}[ht]
\centering
\caption{Ablation Experiment of OGConv Block.}
\label{tab:OGConv}
\resizebox{\columnwidth}{!}{%
\begin{tabular}{l|rrrr|r}

\hline Model & Acc. & Pre. & Rec. & F1. & Params(M) \\ \hline

PCL (Hetero) & 0.9029 & 0.9130 & 0.9037 & 0.9043 & 0.77 \\ \hline
OG-PCL without $K/D$ & 0.9054 & 0.9209 & 0.9065 & 0.9057 & 0.83 \\
OG-PCL & 0.9175 & 0.9312 & 0.9187 & 0.9186 & 0.83 \\ \hline

\end{tabular}}
\end{table}

\section{Conclusion}

In this work, we introduced the Occupancy-Gated Parallel-CNN Bi-LSTM (OG-PCL) to process sparse point cloud signals generated by 3D millimeter-wave radar. Compared with existing point-cloud processing networks, OG-PCL achieves the RadHAR state-of-the-art accuracy of $91.75\%$ on the RadHAR dataset while employing a markedly lighter parameter budget (0.83 M). The tri-view parallel design preserves information across three spatial dimensions at low computational cost, and the Occupancy-Gated Convolution (OGConv) block enhances feature extraction under sparsity through a mask-guided gating mechanism and a compensation factor that stabilizes responses. These properties together render OG-PCL highly suitable for real-time radar-based HAR on lightweight platforms.

Beyond radar-specific benefits, the OGConv component provides a general sparsity-aware mechanism that can be extended to other sparse signal processing tasks in radar and computer vision. While the results validate the effectiveness of our approach, several directions remain for future exploration. First, extending evaluation to larger and more diverse radar datasets would further establish generalization capabilities across environments and subjects. Second, hardware-aware optimizations, including quantization and pruning, could further reduce memory footprint and latency for edge deployments. Finally, investigating end-to-end deployment on target embedded hardware would help translate the method from experiments to real-world applications.




\vfill\pagebreak

\bibliographystyle{IEEEbib}
\bibliography{strings,refs}

\end{document}